\documentclass[fleqn,usenatbib]{mnras}

\usepackage{newtxtext,newtxmath}

\usepackage[T1]{fontenc}
\usepackage{ae,aecompl}

\usepackage{graphicx}	
\usepackage{amsmath}	
\usepackage{amssymb}	
\usepackage{subcaption}
\usepackage[utf8x]{inputenc}
\usepackage{wrapfig}

\newcommand{\Msun}{\,{\rm M}_{\sun}}

\newcommand{\Gyr}{\,{\rm Gyr}}
\newcommand{\kpc}{\,{\rm kpc}}
\newcommand{\FeH}{\lbrack{\rm Fe/H}\rbrack \,}

\title[GCs and stellar streams in E-MOSAICS]{Fossil stellar streams and their globular cluster populations in the E-MOSAICS simulations}

\author[M. E. Hughes et al ]{Meghan E. Hughes,$^{1}$\thanks{E-mail: M.Hughes1@2013.ljmu.ac.uk}
Joel Pfeffer,$^{1}$ 
Marie Martig,$^{1}$ 
Nate Bastian,$^{1}$ 
\newauthor
Robert A. Crain,$^{1}$
J. M. Diederik Kruijssen$^{2}$ 
and Marta Reina-Campos$^{2}$ 
\\
$^{1}$Astrophysics Research Institute, Liverpool John Moores University, 146 Brownlow Hill, Liverpool L3 5RF, UK\\
$^{2}$Astronomisches Rechen-Institut, Zentrum f\"{u}r Astronomie der Universit\"{a}t Heidelberg, M\"{o}nchhofstra{\ss}e 12-14, 69120 Heidelberg, Germany\\
}

\date{Accepted 2018 October 23. Received 2018 October 23; in original form 2018 September 6}

\pubyear{2018}

\begin{document}
\label{firstpage}
\pagerange{\pageref{firstpage}--\pageref{lastpage}}
\hyphenation{kruijs-sen}
\maketitle

\begin{abstract}
Stellar haloes encode a fossil record of a galaxy's accretion history, generally in the form of structures of low surface brightness, such as stellar streams. While their low surface brightness makes it challenging to determine their age, metallicity, kinematics and spatial structure, the infalling galaxies also deposit globular clusters (GCs) in the halo, which are bright and therefore easier to observe and characterise. 
To understand how GCs associated with stellar streams can be used to estimate the stellar mass and the infall time of their parent galaxy, we examine a subset of 15 simulations of galaxies and their star clusters from the E-MOSAICS project. E-MOSAICS is a suite of hydrodynamical simulations incorporating a sub-grid model for GC formation and evolution.
We find that more massive accreted galaxies typically contribute younger and more metal rich GCs. This lower age results from a more extended cluster formation history in more massive galaxies. In addition, at fixed stellar mass, galaxies that are accreted later host younger clusters, because they can continue to form GCs without being subjected to environmental influences for longer.
This explains the large range of ages observed for clusters associated with the Sagittarius dwarf galaxy in the halo of the Milky Way compared to clusters which are thought to have formed in satellites accreted early in the Milky Way's formation history. Using the ages of the GCs associated with the Sagittarius dwarf, we estimate a virial radius crossing lookback time (infall time) of $9.3 \pm 1.8\Gyr$.

\end{abstract}

\begin{keywords}
Galaxy: evolution --- Galaxy: formation --- globular clusters: general --- Galaxy: halo --- Galaxy: stellar content --- galaxies: star formation \vspace{6mm}
\end{keywords}

\section{Introduction}
In the current galaxy formation paradigm, galaxies grow hierarchially through the accretion of diffuse gas and dark matter via filaments and mergers with other galaxies (e.g. \citealt{White1978,White1991}). Mergers with other galaxies can be in the form of a major merger: where two galaxies of similar mass collide; or a minor merger: where a galaxy of lower mass is accreted onto a more massive galaxy. Signatures of both types of mergers can be observed in the local Universe today in the form of substructure in a galaxy's gas, stars and globular cluster (GC) population. Substructure comes in a variety of forms such as shells, streams and planes. 
An abundance of substructure has been observed  in our galaxy, both in the form of overdensities of stars and kinematically \citep{Majewski1996,Newberg2002,Belokurov2006,Starkenburg2009,Martin2014,Shipp2018}, in M31 \citep{Ibata2001,McConnachie2003,Kalirai2006} and other nearby galaxies \citep{Shang1998,Martinez2008,Cohen2014,Merritt2016,Abraham2018}. This work focuses on substructure in the form of stellar streams. \par

Perhaps the most-studied substructure is the Sagittarius stream, which originates from the Sagittarius dwarf galaxy and currently resides in the halo of the Milky Way (MW, \citealt{Ibata1995}). The Sagittarius dwarf galaxy is our closest satellite galaxy with its nucleus just $16 \kpc$ from the Galactic centre \citep{Ibata1995}. It is also the brightest Galactic dwarf spheroidal galaxy and has an estimated current total mass of $\approx 2.5 \times 10^{8} \Msun $ \citep{Law2010}.
Sagittarius is elongated along the direction towards the MW centre which suggests it is undergoing strong tidal distortion before being integrated into our galaxy \citep{Majewski2003}.  The Sagittarius stream is thought to host 7-11 globular clusters and open clusters with high to moderate confidence \citep{Bellazzini2003,Forbes2010,Law2010b},  although the distinction between open clusters and GCs is somewhat arbitrary. Overall, between $25-40 \%$ of the MW's GC population are thought to have been accreted from dwarf galaxies \citep{Forbes2010,Kruijssen2018,Kruijssen2018b}. In M31, there is a striking spatial correlation between stellar substructure and GCs beyond 30$\kpc$ from the galactic centre \citep{Mackey2010}. It was concluded that there is a less than 1\% chance that these GCs are in their spatial configuration by chance \citep{Mackey2010,Veljanoski2014}  and are therefore likely to have been accreted with the stars comprising the substructure.\par
 
It has been postulated that substructures in a galaxy's halo will present different stellar ages and metallicities than the bulk of the stellar halo because of their late infall onto the central galaxy and their smaller stellar mass \citep{Ferguson2002,Johnston2008}.  Therefore, we might also expect tangible differences between the halo population of GCs and those which are associated with stellar streams. GCs associated with stellar streams, by construction, formed in a galaxy with a different star formation history, and hence a different GC formation history, than the galaxy in which they currently reside. Therefore stars and GCs associated with a particular stellar stream are expected to exhibit a different age-metallicity relationship to those formed in the central galaxy \citep{Forbes2010,Dotter2011,Leaman2013,Kruijssen2018}.
\cite{Mackey2013} estimate that 2 of the GCs (PA-7 and PA-8) associated with the M31 substructure known as the South West Cloud have ages of 6-10 \Gyr, which makes them at least 3 $\Gyr$ younger than the oldest MW GCs.
However, there is no evidence that GCs associated with stellar streams are in general younger than the rest of the GC population. In fact, some GCs associated with the Sagittarius stream are classified as old halo clusters \citep{Mackey2005}, and from proper motion estimates of MW GCs it has also been suggested that young clusters are also formed in-situ \citep{Sohn2018}. 
 \par

GCs form in tandem with the field stars comprising galaxies \citep{reinacampos18b}, taking part in merger events alongside their parent galaxies. With photometry and regular spectroscopy it is difficult to find stars from a tidally disrupted galaxy, therefore the greater surface brightness of its associated GCs renders them more readily identifiable against the background of field stars. This makes a galaxy's GC population a powerful means of inferring a picture of its formation \citep[e.g.][]{Harris1991,Forbes1997,Brodie2006,Kruijssen2018,Kruijssen2018b}. \par

This work uses simulations from the E-MOSAICS project \citep{Pfeffer2017,Kruijssen2018} to investigate properties of the GCs associated with stellar streams at $z=0$. It is organised as follows; in Section \ref{2} we give an overview of the simulations used for this work and then in Section \ref{3} we discuss how we identify stellar streams in the simulations. In Section \ref{4} we examine the ages and the metallicities of the GCs associated with stellar streams relative to those of other GCs associated with the host galaxy and relate these properties to the GC parent galaxy mass and infall time. In Section \ref{5} we investigate the relationship between the GC formation history, galaxy mass and infall time to provide a method to estimate the infall time of the Sagittarius dwarf galaxy and in Section \ref{6} we compare the results in this paper to observables.

\section{Simulations \label{2}}
For the purpose of this work we use the E-MOSAICS (MOdelling Star cluster population Assembly in Cosmological Simulations within EAGLE, \citealt{Pfeffer2017,Kruijssen2018}) suite of simulations which follow the co-formation and evolution of galaxies and their GC\footnote{In this work we consider a GC to be any star cluster  that is above $2 \times 10^4 \Msun$.} populations in a cosmological context. This is achieved by combining the MOSAICS \citep{Kruijssen2011} sub-grid model of stellar cluster formation and evolution into the software used to conduct the EAGLE (Evolution and Assembly of GaLaxies and their Environments, \citealt{SchayeEAGLE2014,Crain2015}) galaxy formation simulations as described in \citet{Pfeffer2017} and \citet{Kruijssen2018}.\par

EAGLE is a set of  hydrodynamical simulations of the formation of a cosmologically representative sample of galaxies in a $\Lambda$CDM cosmogony. The simulations use a heavily-modified version of the smoothed particle hydrodynamics (SPH) code GADGET3 (last described by \citealt{Springel2005}). The main modifications are to the hydrodynamics algorithm, the time-stepping criteria (see \citealt{SchayeEAGLE2014} for more detail), and the addition of a suite of sub-grid models which govern processes acting on scales below the simulation's numerical resolution. \cite{Schaller2015} investigates the impacts of these modifications on the EAGLE galaxy population.
The routines include sub-grid radiative cooling \citep{Wiersma2009}, star formation \citep{Schaye2008}, stellar feedback \citep{DallaVecchia2012}, chemical evolution \citep{Wiersma2009}, gas accretion onto, and mergers of,  super massive black holes (BHs) \citep{Rosas-Guevara2015,SchayeEAGLE2014} and active galactic nuclei (AGN) feedback \citep{Booth2009,SchayeEAGLE2014}. The efficiency of the stellar feedback and the BH accretion is included in the simulation calibration to match the $z=0$ galaxy stellar mass function and the sizes of disc galaxies, and the AGN feedback is calibrated to produce the known relationship between the BH mass and the galaxy stellar mass. The standard resolution EAGLE simulations yield a galaxy stellar mass function that reproduces the observed function to within 0.2 dex over the well-sampled and well-resolved mass range. The simulations also reproduce other observables, such as the galaxy specific star formation rates and the total stellar mass of galaxy clusters. For a full description of the models, see \cite{SchayeEAGLE2014}.
To follow the formation of a galaxy halo, the SUBFIND algorithm \citep{Springel2001,Dolag2009} is used to identify subhaloes (galaxies) in the simulations, from which we construct merger trees using the method described by \cite{Pfeffer2017}.\par

Modelling star cluster systems requires treatment of the star cluster formation, evolution and disruption processes. E-MOSAICS adopts a star cluster formation model based on observations of young star clusters, under the assumption that young star clusters, GCs and open clusters have a common formation mechanism \citep{Longmore2014,kruijssen15b,Bastian2016}. Whenever a stellar particle is formed, some fraction of the stellar mass is considered to reside in bound clusters, with identical age and metallicity as the parent stellar particle.
Cluster formation is regulated by a cluster formation efficiency (CFE, \citealt{Bastian2008}), i.e. the fraction of all star formation across the galaxy which occurs in bound clusters, which increases with star formation rate surface density \citep[e.g.][]{adamo15b}. E-MOSAICS adopts the environmentally dependent description of the CFE from the \citet{Kruijssen2012MNRAS} model, which relates the CFE to the properties of the interstellar medium (ISM), reproducing the observed trend. Secondly, we consider an environmentally dependent initial cluster mass function as in the model of \citet{Reina-Campos2017}, which relates the maximum mass of the molecular cloud to the Toomre mass \citep[as in][]{kruijssen14c} and also includes the effects of stellar feedback. The mass loss and potential disruption of clusters is also followed. Mass loss occurs via stellar evolution (which is calculated for each stellar particle by the EAGLE model,\citealt{Wiersma2009}) and dynamical evolution in the form of two-body relaxation, tidal shocks and dynamical friction \citep{Kruijssen2011,Pfeffer2017}. Clusters are evolved down to mass of 100 $\Msun$ after which they are assumed to be completely disrupted. \par

E-MOSAICS predicts the properties of the young star clusters in the simulated galaxies which are in good agreement with observations of young clusters in nearby disc galaxies (Pfeffer et. al in prep.). The range in the number of GCs is consistent with observed ones in the MW and M31. This is discussed in more detail in \citet{Kruijssen2018b}, where we explicitly compare the number, metallicity distributions, and spatial density profiles of the populations to the observed values of the MW and M31. The radial distribution of the birth pressure of the clusters matches that of the observations of \citet{leroy2008} \citep{Pfeffer2017}. The CFE radial distribution is similar the observed distributions of \citet{Silva-Villa2013,johnson2016} and the global CFE at z=0 of all the galaxies shows the same range as that observed \citep[1-50\%, e.g.][]{adamo15b,johnson2016} \citep{Pfeffer2017}. In addition, E-MOSAICS reproduces the properties of GC populations. One such property is the blue tilt \citep{Usher2018}, where there is a lack of massive metal poor GCs, first observed by \citet{Harris2006}. The ages of the GCs in the E-MOSAICS simulations reproduce those of observed systems, for example \citet{reinacampos18b} show that not only are the median ages of MW and extragalactic GCs reproduced, but also the observed age offset between metal-poor and metal-rich GCs \citep[e.g.][]{Brodie2006,Forbes2015}. In addition, the E-MOSAICS galaxies are consistent with the specific frequency, spatial distribution and upper end ($> 10^{5} \Msun$) of the mass function of GCs in the Milky Way \citep{Kruijssen2018}
Although many properties of GC populations are reproduced, the number density of low mass clusters in E-MOSAICS is over predicted. This is due to the lack of a cold, dense gas phase in the EAGLE model, which would disrupt many of these clusters (as discussed in \citealt{Pfeffer2017}) and it will be addressed in a future generation of models. In this work, the over-prediction is minimised by focusing on clusters with masses $M>2\times 10^{4} \Msun$ at $z=0$. This still leaves an overabundance of low-mass clusters relative to the GC populations observed in massive galaxies, but for the dwarf galaxies we are studying in this work, this may not be a significant problem since there is tentative evidence for dwarf galaxies having an excess of lower-mass GCs \citep{Huxor2014}. 

E-MOSAICS is currently a suite of 25 zoom-in simulations of MW-mass galaxies. They are selected solely on the basis of their halo mass, meaning they span a wide range of formation histories. This makes the E-MOSAICS galaxies well suited to investigate the properties of GCs associated with stellar streams in a range of environments. We want to only include galaxies with a disc-like morphology, i.e. somewhat similar to the MW, so we exclude any which have undergone a major merger (a merger with a stellar mass ratio greater than $1/4$) at $z\approx 0$  or are in the process of undergoing a major merger, since this would greatly disrupt the present day configuration of star particles. We also exclude galaxies which are not of disc-like morphology or do not contain any stellar streams. Therefore we finally have a set of 15 zoom simulations of MW-like galaxies which contain streams to carry out our analysis (these are MW01, MW02, MW03, MW05, MW06, MW07, MW08, MW09, MW10, MW12, MW13, MW17, MW20, MW23 and MW24 in Table 1 of \citealt{Kruijssen2018}).\par 

\section{Identifying stellar streams and their associated GCs in E-MOSAICS\label{3}}
\subsection{Stellar stream identification \label{sec3.1}}
\begin{figure*}
  \centering
	\includegraphics[width=\linewidth]{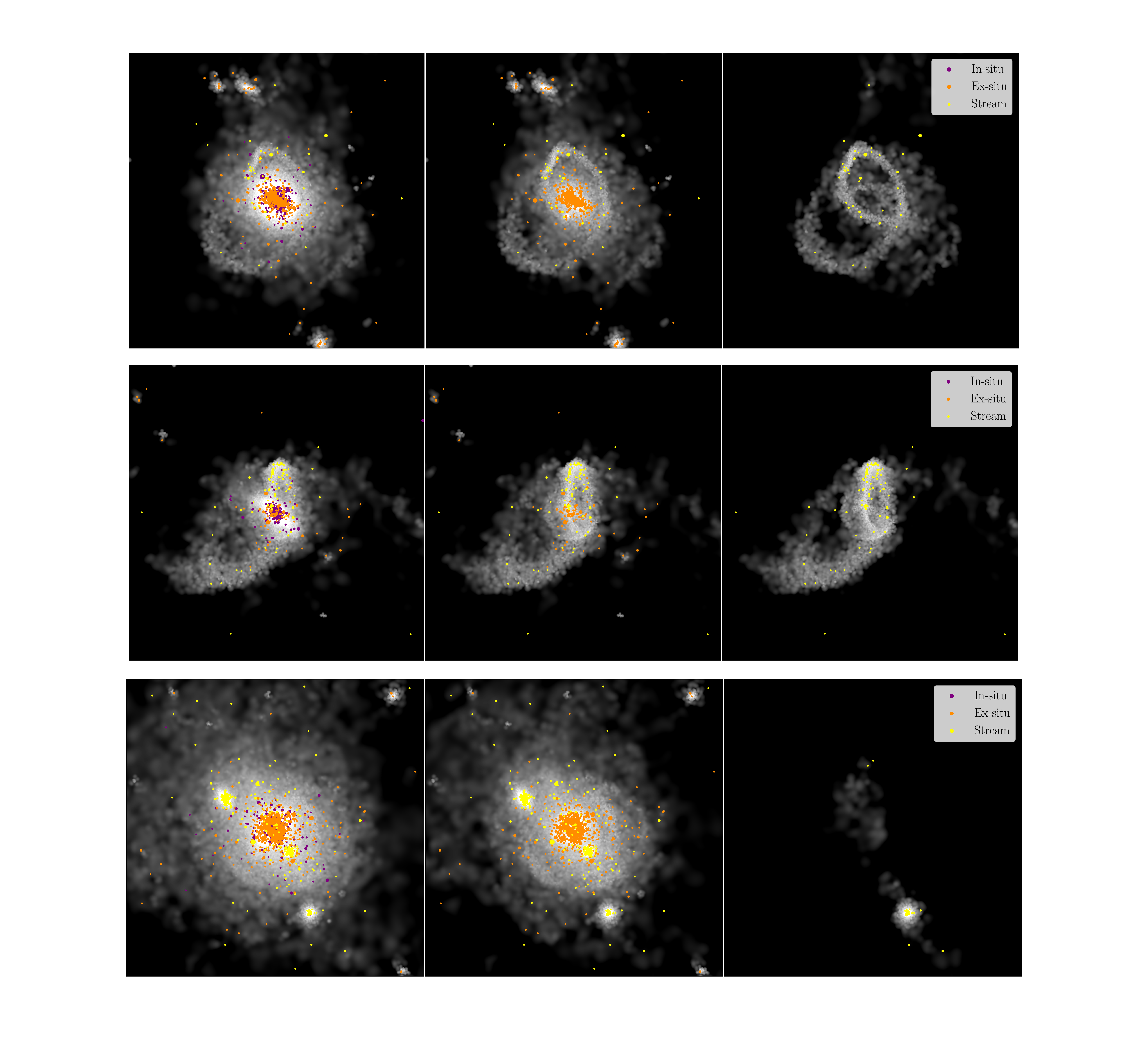}
     \caption{Stellar density plots of 3 of the haloes which show clear streams. From top to bottom, we show galaxies MW03, MW09 and MW17. From left to right, the plots show the main galaxy, the accreted component (everything that did not form in the central galaxy) and one clear stream. Each panel is 200 kpc on a side.}
      \label{fig:full_galaxies}

\end{figure*}

\begin{figure*}
      \begin{subfigure}[b]{7cm}
        \centering
    	\includegraphics[width=\linewidth]{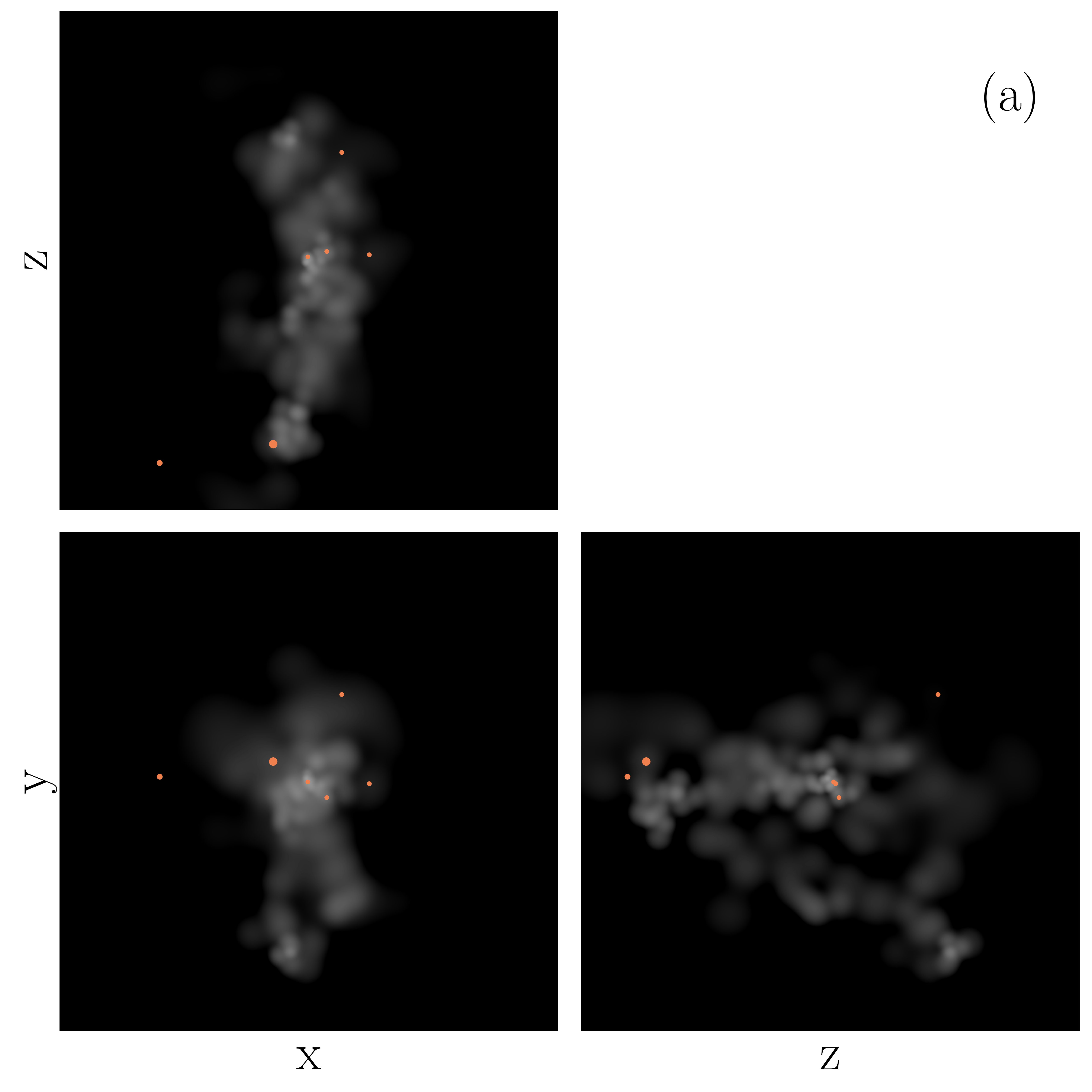}
                \label{fig:stream2}

      \end{subfigure}
      \begin{subfigure}[b]{7cm}
        \centering
	\includegraphics[width=\linewidth]{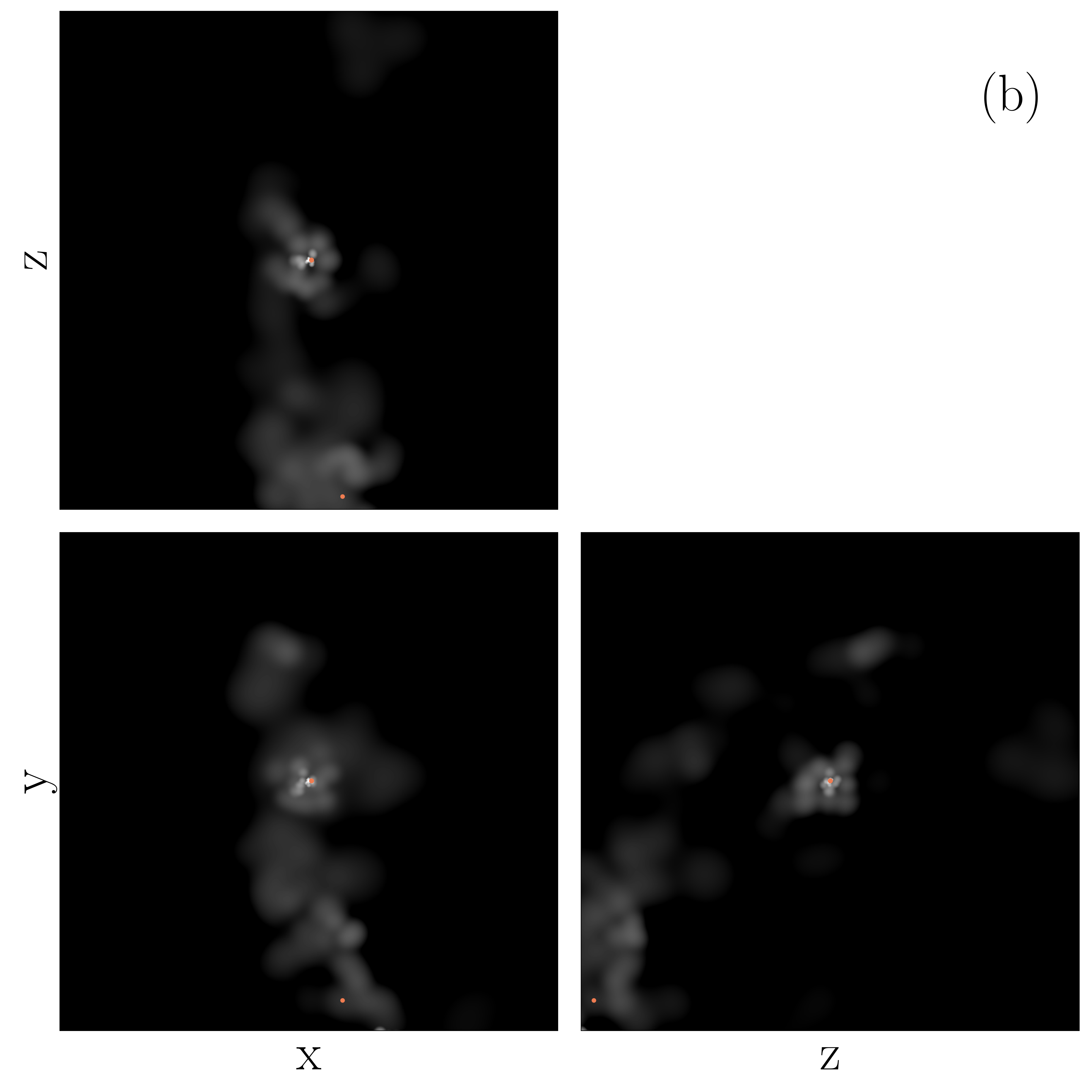}
            \label{fig:stream3}

      \end{subfigure}
      \begin{subfigure}[b]{7cm}
        \centering
    	\includegraphics[width=\linewidth]{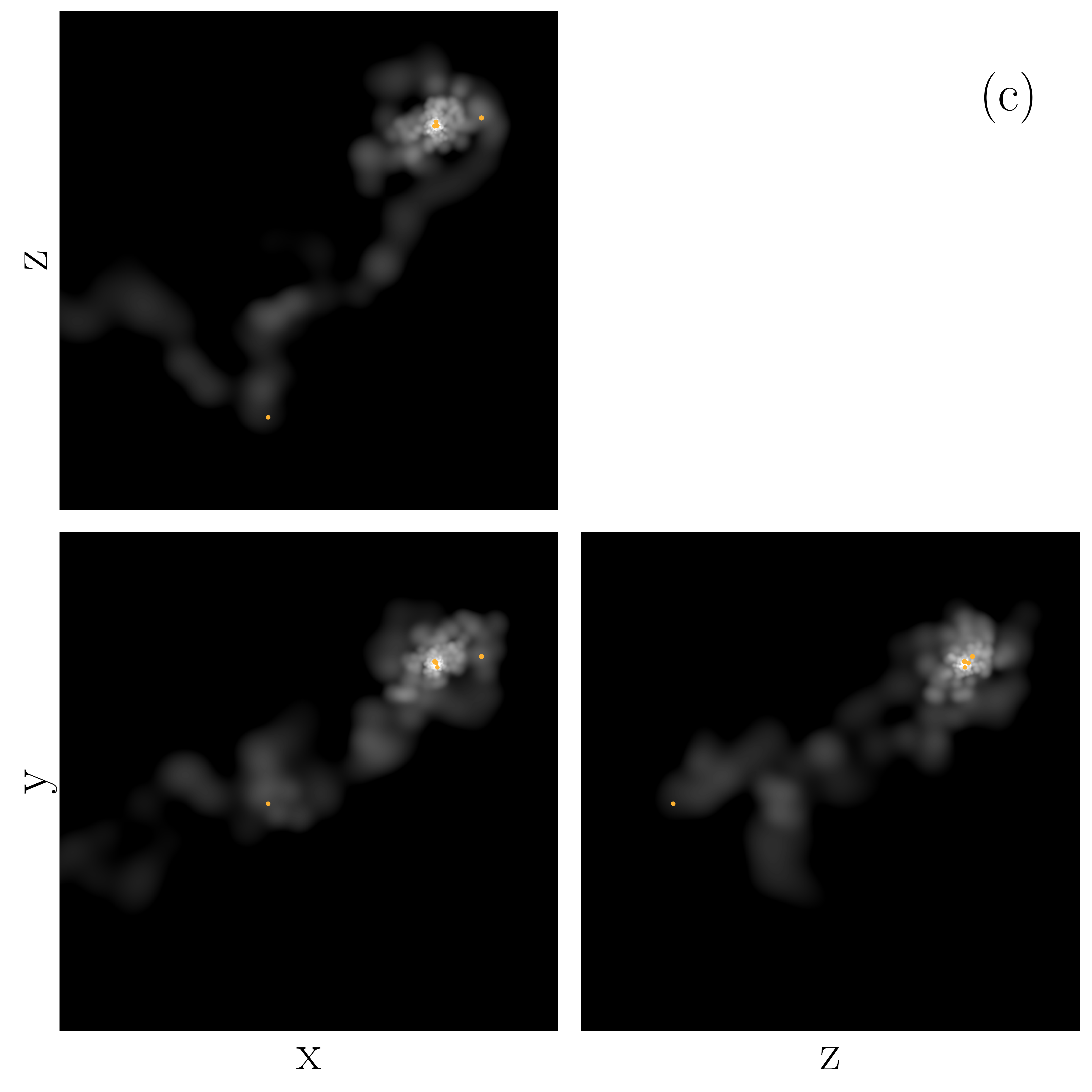}
                \label{fig:stream4}
    
          \end{subfigure}
      \begin{subfigure}[b]{7cm}
        \centering
	\includegraphics[width=\linewidth]{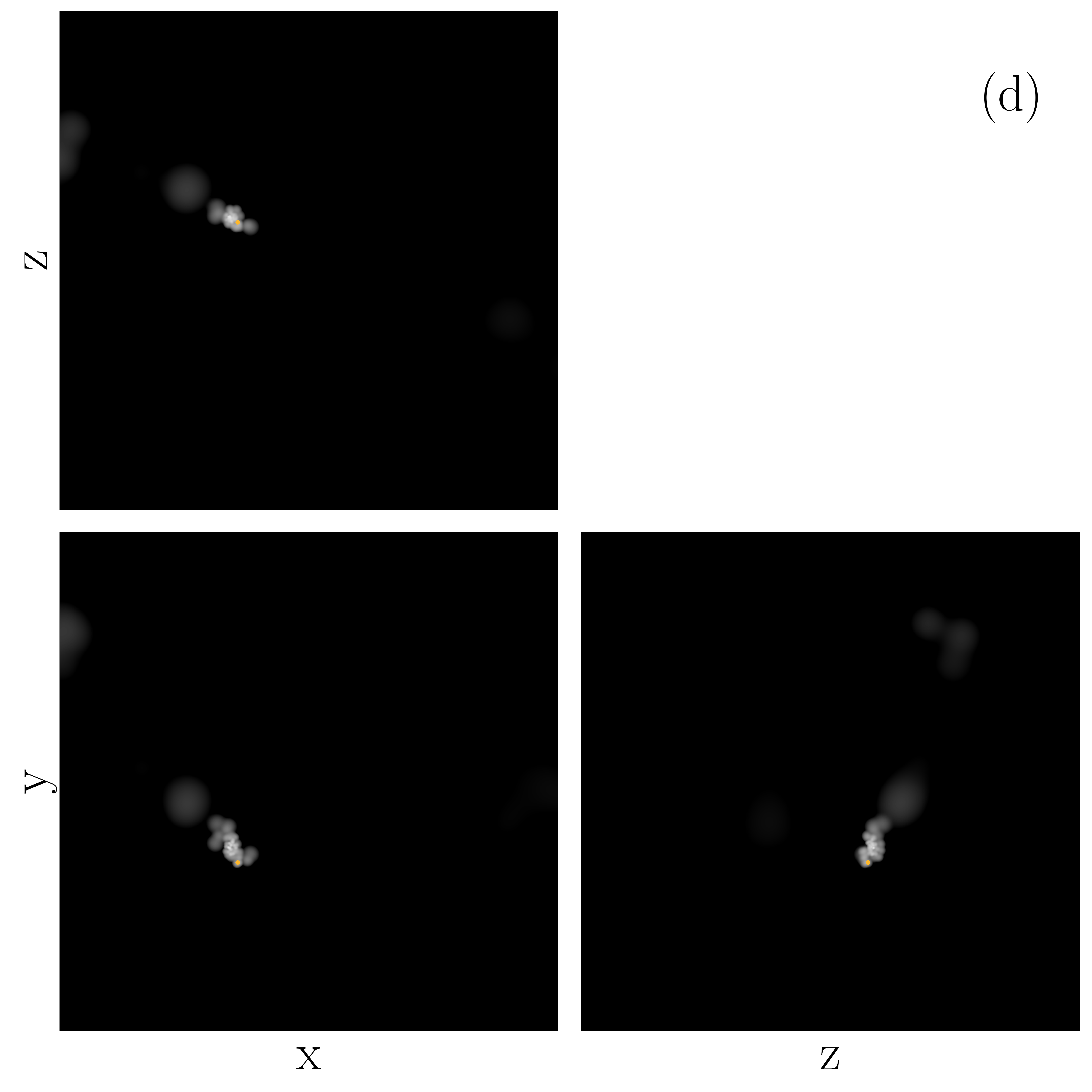}
            \label{fig:stream5}
            
      \end{subfigure} 
      \caption{Examples of stellar substructure generated during the accretion of a single galaxy. These panels highlight the high diversity in the classification of a stellar stream. All of these events were placed into the stream category. From panel (a)--(d) these are accretion events from MW12, MW17, MW05 and MW13.}
      \label{fig:all_streams}
\end{figure*}

The first step towards being able to describe the GC population in stellar streams requires the identification of such structures and their associated GCs in our suite of simulations. The following describes the method we implemented.\par

The simulations record the history of particles, enabling us to trace star particles and their associated GCs from formation until $z=0$ . This means we can assign a parent galaxy to the stellar particles and GCs. If their parent galaxy is not the main galaxy then they must have been accreted onto the main galaxy via a merger.
This allows us to view the current positions of the stars and GCs associated with each individual accretion event throughout the main galaxy's formation history, without the contamination from any other stars or GCs in the main galaxy. From this, we can determine whether or not the stars are in a stream like configuration.\par

The $z=0$ positions of the stars from each accreted galaxy that contain more than 100 star particles at $z=0$, corresponding to a stellar mass of $\approx 10^{7} \Msun$,  is shown in a stellar density map in three projections.
 Fig. \ref{fig:full_galaxies} shows three of the galaxies with clear stellar streams (MW03, MW09 and MW17 from top to bottom). The coloured points in this plot represent all of the GCs with a mass greater than $2 \times 10^{4} \Msun$. The left hand panels show all the stars and GCs in the main galaxy, the middle panels show just the accreted stars and GCs and the right hand panels show the stars and GCs from just one of the accreted galaxies whose current configuration is classified as stream-like.
All of these figures show structures which are unambiguously classified as stellar streams. \par

Fig. \ref{fig:all_streams} shows the stellar density map in three projections of stream-like accretion events, it is from these three 2-D projections of individual accretion events that we identify stream-like substructures. Fig \ref{fig:all_streams} illustrates that we find a considerable diversity of substructures. This makes categorising the accretion events difficult in a minority of cases. In order to combat this, four of the authors of this paper partook in the classification of streams. A universal classification method was developed for all authors to follow. For the event to be classified as stream-like the stellar density had to be elongated in at least two of the projections. The identification of streams is complicated by the presence of gravitationally bound, spheroidal relics of accreting satellites, as well as shell-like structures. If the bound object is considered to have a significant tail-like structure then it is classified as a stream. Shell-like structures are more difficult to categorise and therefore we exclude them from the stream sample. Over the 15 galaxies, 3-7 streams are identified per halo with a mean number of streams per halo of 4.5. The percentage of accreted galaxies with a mass greater than $10^7 \Msun$ which leave streams varies between 14-36 \% with a mean of 21.4\%. \par
 
Furthermore, all GCs formed in a galaxy generating a stellar stream were included in the `on stream’ category, regardless of their projection onto the stream. Therefore, any GCs that formed in the accreted galaxy, but are not currently visually associated with it, have been included regardless. This is done to account for observers potentially having chemo-kinematic information about the GCs. For example, Palomar 12 is thought to have once been associated with the Sagittarius dwarf galaxy, yet it now sits at a wide separation from Sagittarius on the sky \citep{Cohen2004,Sohn2018}. There are also a handful of other GCs which are candidates for once being related to the Sagittarius dwarf even though they are no longer spatially associated with the stellar component \citep{Forbes2010}. \par

Once all of the GCs have been classified either into the stream or non-stream category, the analysis includes a selection on GC properties. First of all, a lower mass limit of  $2 \times 10^{4} \Msun$  is imposed to alleviate the under disruption of low mass clusters in E-MOSAICS (the importance of this is discussed in section \ref{2}). This mass corresponds to a luminosity of $M_{V}\approx -5 $ at old ages ($ > 10 \Gyr$). The PAndAS survey begins to suffer from incompleteness at $M_{V}\approx -6 $ and is $50\%$ complete at $M_{V}\approx -4.1$ \citep{Huxor2014}, therefore this mass limit is reasonable for comparison with the MW and M31.
A radius cut of $\rm{R} >10 \kpc $  from the main galaxy's centre is imposed on all GCs to excise most of the disc GC population. When observing an external galaxy, the central substructure is lost due to the high surface brightness of the main galaxy. This also makes finding GCs in this central region difficult. The radius cut also helps to alleviate the underestimated disruption rate in the centre of the galaxy due to the lack of cold interstellar medium in E-MOSAICS, as discussed in Section \ref{2}.\par

\subsection{Definitions}
We now define several terms that will be used frequently throughout the rest of this paper. In-situ and ex-situ define whether the GCs were formed in the main galaxy or not. This is defined as where the gas particle was prior to forming a stellar particle. Figure 10 in \citet{Pfeffer2017} and figure 5 in \citet{Kruijssen2018} show examples of the merger trees. In these figures, the main branch is highlighted by the thick black line and represents the evolution of the central galaxy. If the gas particle is in a subhalo on the main branch of the merger tree before it becomes a star/cluster population, then this is in-situ star/cluster formation, whereas if the gas particle is on a different branch of the merger tree, then it is ex-situ star/cluster formation. 
We define GCs that are formed whilst bound to the central galaxy but from the gas that has been accreted as in-situ. This may affect a minority of cases where a GC forms just after the satellite galaxy has merged with the main galaxy and the gas particle gets assigned to the main galaxy instead of the satellite. \par

The GCs that are referred to as `stream' are ex-situ GCs by definition, because they had to be formed in a halo other than the main galaxy to be accreted along with the stellar component that then forms a stream. Non-stream GCs are a combination of both in-situ and ex-situ, because they are simply defined as the GCs which are not associated with a stream at $z=0$. \par

For reference, properties which are named in the form ${X}_\mathrm{c}$ refer to the GC properties. More specifically, ${X}_\mathrm{c, stream}$ relates to the median of this particular property on this particular stream and ${X}_\mathrm{c, non-stream}$ relates to the median of this particular property of all the GCs, not including those on the stream in question, but still including those from other streams.  Properties which are named in the form ${X}_\mathrm{sat}$ refer to the accreted galaxy properties. The properties of the GCs we consider are the metallicity ($\FeH$) and the age. The properties of the accreted galaxies considered are the stellar mass (${M_\mathrm{sat}}$) and the infall time ($T_\mathrm{infall}$). The infall time is defined as the last time the galaxy enters the halo of the main galaxy\footnote{Some galaxies undergo multiple crossings of the virial radius.} and is measured in terms of lookback time. The mass of the stream progenitor galaxy is measured when the stellar mass is at a maximum, before the galaxy is affected by tidal stripping.\par

\section{GCs in stellar streams \label{4}}
\subsection{Properties of GCs associated with stellar streams}

\begin{figure}
	\includegraphics[width=\columnwidth]{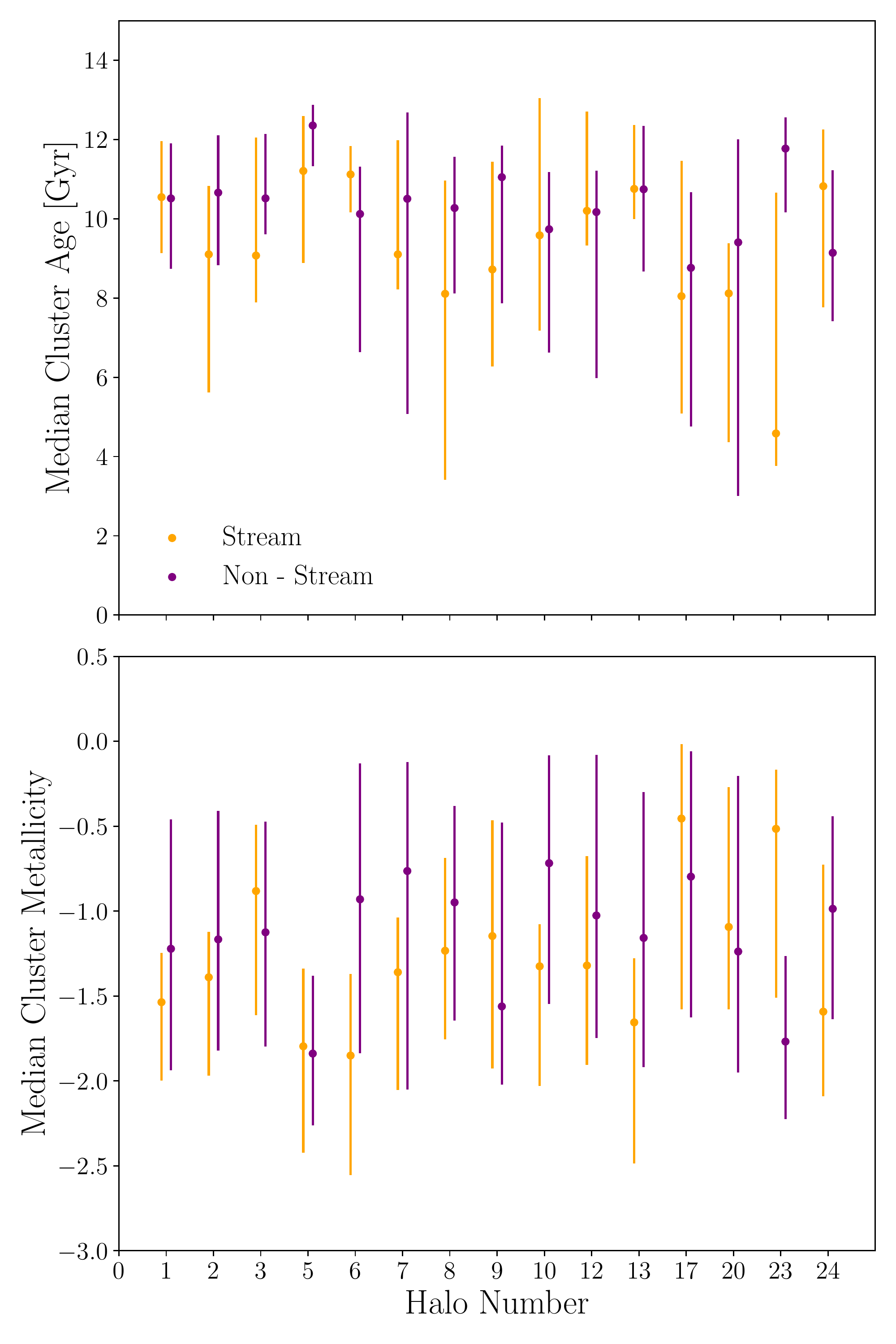}
    \caption{Median ages and metallicities of GCs on and off the streams shown with their 16th and 84th percentile bars. Each pair of points represents one simulated halo, where `stream' refers to the median of all the GCs which are associated with all of the streams in a given halo, and `non-stream' refers to the median of all the GCs which are not associated with a stream in this halo. The GCs have undergone the mass, age and radius cuts mentioned previously. Note the large variation from halo to halo.}
    \label{fig:cluster_properties.pdf}
\end{figure}

We first examine the median ages and metallicities of the GCs on all streams, and GCs not on streams, for each halo.  Fig. \ref{fig:cluster_properties.pdf} shows these median ages and metallicities for the 15 haloes along with their 16th and 84th percentile bars. The GCs associated with streams exhibit diverse properties. The mean difference in the ages of the stream and non-stream populations is $-1.19 \Gyr$  i.e. stream GCs are typically younger) with a standard deviation of $2.15 \Gyr$. The mean difference in the metallicity of the stream and non-stream population is $-0.17$ dex (i.e. stream GCs are typically less metal-rich) with a standard deviation of $0.53$ dex. This diversity motivates a closer scrutiny of the progenitors of the streams. \par

\begin{figure}
	\includegraphics[width=\columnwidth]{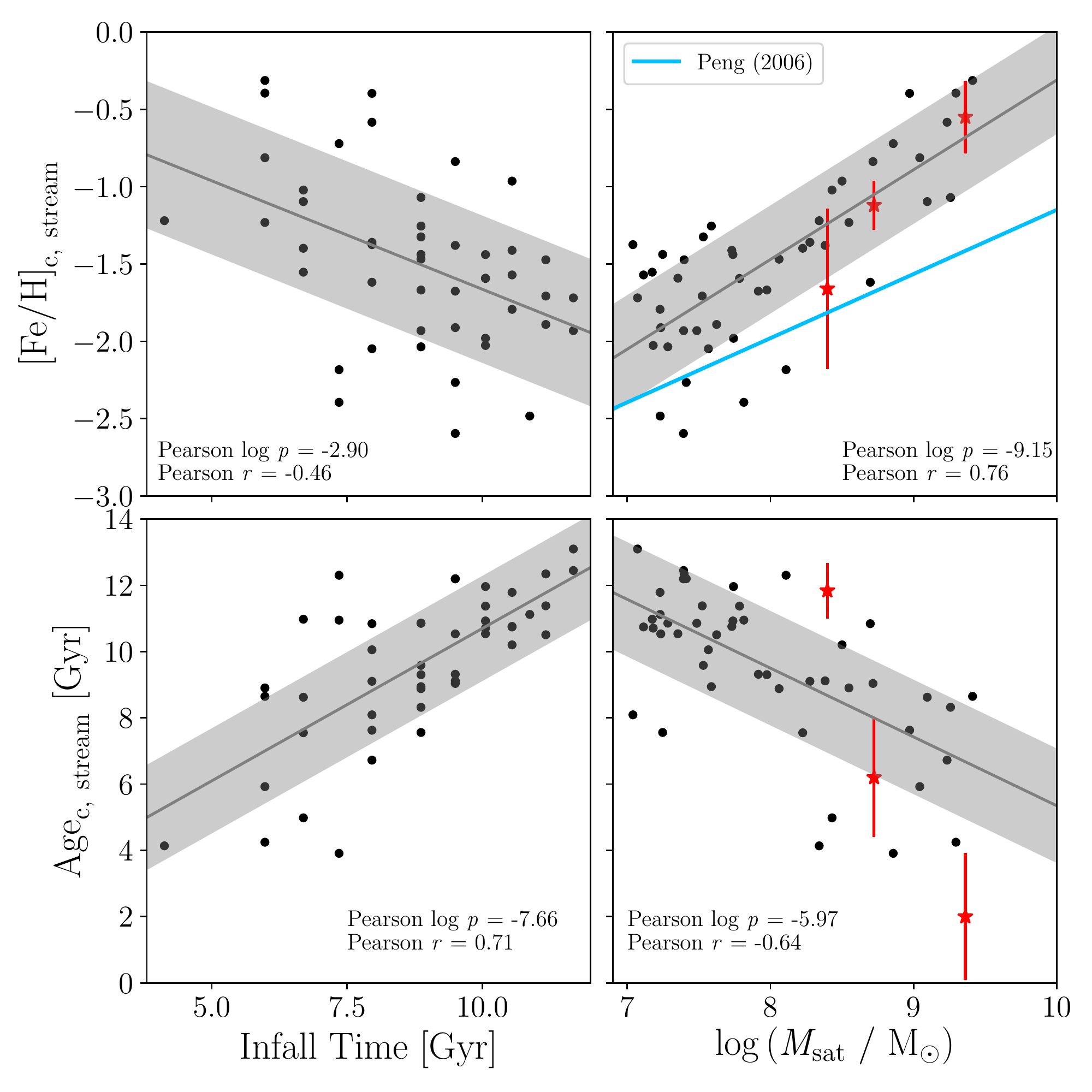}
    \caption{Host galaxy properties (lookback time of the crossing of the virial radius, i.e. the 'infall time', and stellar mass) are plotted against the GC properties (median metallicity and median age) in order to highlight key trends. Here, each point represents an individual stream progenitor galaxy across all simulated haloes. The black lines represent the fit and the grey band represents the $1 \sigma$ scatter of the data around the fit. The red stars represent where the Sagittarius dwarf, the SMC and the LMC (from low to high mass) lie in this parameter space -- see the text for age and metallicity references. The mass-metallicity relation of \citet{Peng2006} for all GCs is shown by the light blue line in the top right panel- this has been extrapolated below stellar masses of $5 \times 10^8 \Msun$. Age and metallicity show a clear dependence on the parent galaxy properties, indicated by the Pearson coefficients quoted in each panel.  }
    \label{fig:virialradius_met_streams.pdf}
\end{figure}

 Fig. \ref{fig:virialradius_met_streams.pdf} shows the relationship between infall time and stellar mass of the stream progenitor galaxies and the median age and metallicity of the GCs they bring into the main halo. The median age of GCs on streams increases with the satellite infall time and decreases with galaxy mass. The median metallicity of GCs on streams decreases with galaxy infall time and increases with galaxy mass. The Pearson $\mathit{r}$ and $\mathit{p}$ coefficients are shown for each of the panels and all the panels show reasonably strong trends. The strongest of these trends is between GC metallicity and galaxy mass (Fig. \ref{fig:virialradius_met_streams.pdf}, top right). \citet{Peng2006} also investigate the relation between GC metallicity and galaxy mass for the GCs in 100 early type galaxies. The relation of \citet{Peng2006} (their figure 13) for all GCs is shown in this panel by the blue line and we find that our relation is steeper than theirs. The shallower relation of \citet{Peng2006} is potentially caused because they study galaxies which are in a cluster environment, whereas the galaxies we are using for this work occupy less dense environments and we resolve much lower galaxy and GC masses. Galaxies which reside in cluster environments are likely to have been quenched and therefore dwarf galaxies around MW-like galaxies have more extended star formation histories and therefore contain higher metallicity clusters. Also, galaxies in clusters are more likely to grow via the accretion lower mass galaxies which bring with them lower metallicity GCs. Our steeper relation could also be a simulation effect in that we do not disrupt enough higher metallicity clusters, although this is partially ruled out by confirming that the local group dwarf galaxies lie within our steeper relation.
 
 We compare our results with observations by placing the Sagittarius dwarf, the SMC and the LMC in this figure. We take the GCs most likely to be associated with the Sagittarius stream from \cite{Law2010b} and find a median \FeH $= -1.5$  and a median age of $11.84 \Gyr$(using the average ages and metallicities from \citealt{Forbes2010,Dotter2010,Dotter2011} and \citealt{VandenBerg2013})\footnote{We have excluded Berkley 29 and Whiting 1 from this analysis to be consistent with our mass cut.}.  The LMC and SMC are also currently falling into the halo of the MW and are beginning produce a stellar stream-like structure (e.g. \citealt{DOnghia2016}). If we were to plot the LMC and SMC GCs  on this plot with median GC metallicity and age of $\FeH= -0.55$ and $2 \Gyr$ \citep{Suntzeff1992,Gilmozzi1994,Hunter1995,Costa1998,Olsen1998,Dirsch2000,Hill2000,Geisler2003,Piatti2003,Mackey2004,Mackey2007,Ferraro2006,Kerber2007,Mucciarelli2008,Mucciarelli2009,Mucciarelli2011,Muciarelli2012,Palma2013,Li2013,Mackey2013,Mucciarelli2014,Wagner-Kaiser2017} and $\FeH =-1.12$ and $6.2 \Gyr$\citep{dacosta1998,Sirianni2002,Glatt2008,Dalessandro2016}, respectively.
 With a stellar mass of  $\approx (2-3) \times 10^{8}  \Msun$ \citep{Niederste-Ostholt2010} this would place Sagittarius slightly lower than our relation in the top right panel but still within the scatter. The SMC and LMC have  masses of $2.3 \times 10^{9} \Msun$ and $5.3 \times 10^{8} \Msun$, respectively \citep{James2011}, they also lie within our mass-metallicity relation for satellites of late type galaxies. We can also place Sagittarius, the SMC and the LMC in the bottom right panel. The LMC lies much lower than our relation here. However, the black points in this figure represent satellite galaxies which are now streams, and the SMC and the LMC have not yet formed a stream like structure owing to their relatively recent accretion. The comparison here with the progenitors of streams in the simulations may therefore may not be wholly like-for-like. \par 
 There is a wide range in the properties of the stellar streams shown in Fig. \ref{fig:virialradius_met_streams.pdf}, causing the large scatter in the global GC properties of each halo as shown in Fig. \ref{fig:cluster_properties.pdf}. Streams with more massive progenitors contain younger and more metal-rich GCs than streams with less massive progenitors. Streams that fell into the main galaxy more recently also have younger and more metal-rich GCs..
In the following sections we investigate mass and the infall time of the galaxies and the properties of their GCs.\par

Finally, note that the infall time is discreet due to the snapshot resolution of the simulations.
\subsection{Comparisons of GC properties on and off streams \label{4.2}}
\begin{figure}
	\includegraphics[width=\columnwidth]{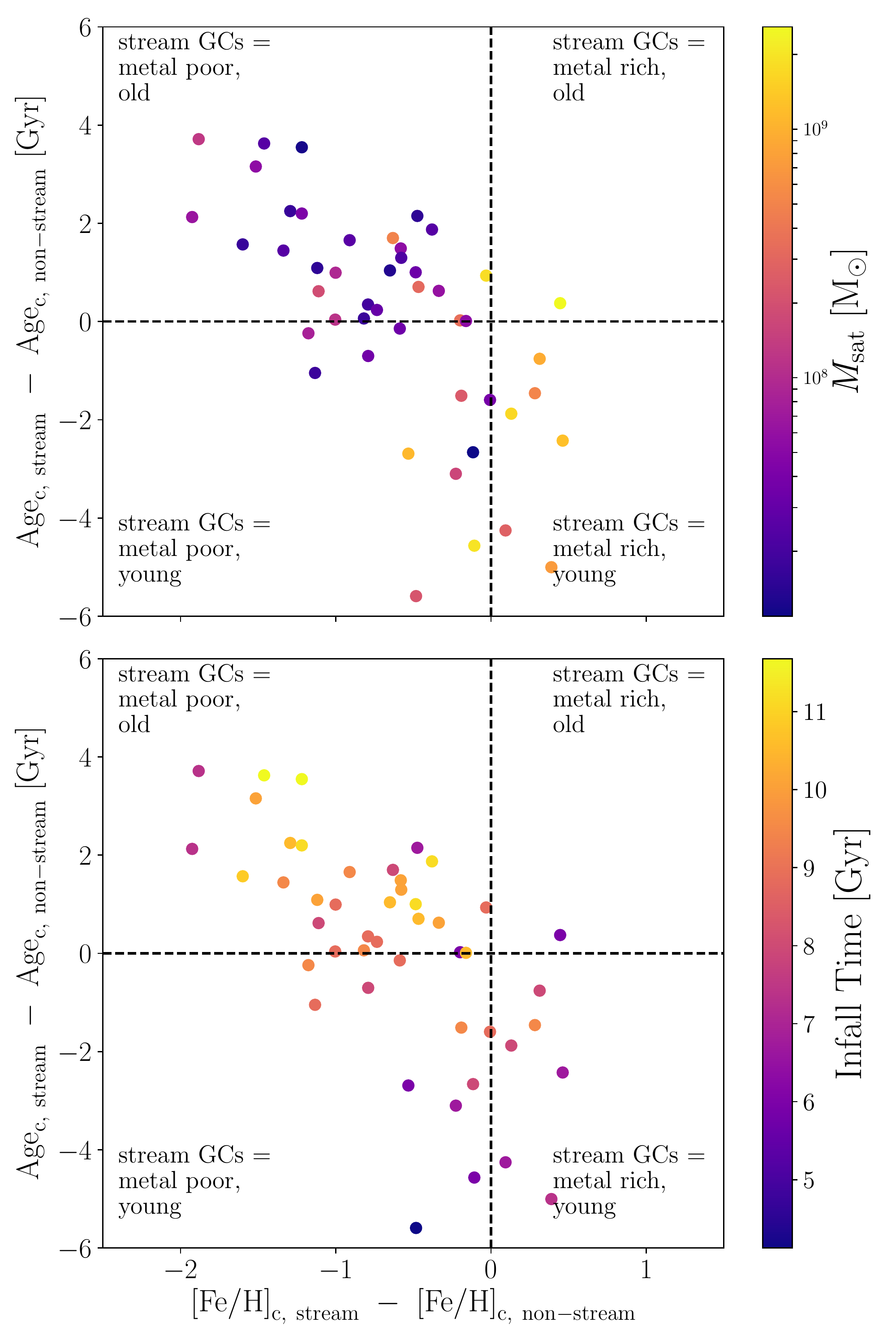}
    \caption{The difference in the median GC ages between the stream and non stream GCs, within the same halo, plotted as a function of the difference in their median metallicities. Each point represents one stream. The non-stream population refers to all the GCs which survive the various property cuts which do not lie on this particular stream; it therefore includes GCs which lie on other streams in this halo, GCs which have been accreted but do not lie on a stream and GCs formed in the main galaxy. Top panel: the colours represent the host galaxy's stellar mass. Bottom panel: the colours represent the virial radius crossing time. There is an anti correlation between age and metallicity.  More massive galaxies which crossed the virial radius more recently are more likely to have GCs on streams which are younger and more metal rich. }
    \label{fig:met_age.pdf}
\end{figure}

We now investigate the properties of the GCs on one particular stream relative to the rest of the GC population (the GCs not associated with this particular stream), and then, relate it to the mass and infall time of the stream progenitor galaxy. The motivation for this investigation is that in some observational cases we may be able to associate a given set of GCs with a stellar stream but do not know where the rest of the GCs in the halo came from. In Fig. \ref{fig:met_age.pdf} each point represents a single stream. The $\mathit{x}$-axis represents the median $\FeH$ of the GCs on the stream relative to the median $\FeH$ of the rest of the population. The $\mathit{y}$-axis represents the median age of the GCs on the stream relative to the median age of the rest of the GC population. The points in the top panel are coloured by the maximum stellar mass of the satellite galaxy before infall and the colours in the bottom panel represent the infall lookback time of the stream progenitor galaxy. Streams that have younger GCs also have more metal rich GCs and come from more massive progenitor stream galaxies that are accreted later.  It is these two competing effects that cause the variation among galaxies we see in Fig. \ref{fig:cluster_properties.pdf}.\par

Using  Fig. \ref{fig:met_age.pdf}, we can restrict the sample to only the most massive streams that fell into the halo recently, since these are those that are readily observable. These streams present younger and more metal rich GCs than the rest of the population. This can also be seen in M31, where the observable streams do show younger GCs (Mackey et al. in prep.). Lower mass streams that fell into the halo of the main galaxy longer ago tend to harbour GCs that are older and more metal poor than the rest of the population.\par

Note the lack of GCs in the top right quadrant of Fig. \ref{fig:met_age.pdf}: there are no satellite galaxies that bring with them relatively old and metal-rich GCs.  In order to populate this region of the plot, the GC host galaxy would have had to self-enrich faster than the present day central galaxy. But the enrichment history and metallicity depends on galaxy mass \citep{Petropoulou2012}, so for a galaxy which forms a stream in the halo of the main galaxy, this is unlikely.\par

\section{The Relationship Between GC Formation History, Galaxy Mass and Infall Time \label{5}}

\subsection{Total age range of GCs \label{5.1}}

\begin{figure}
	\includegraphics[width=\columnwidth]{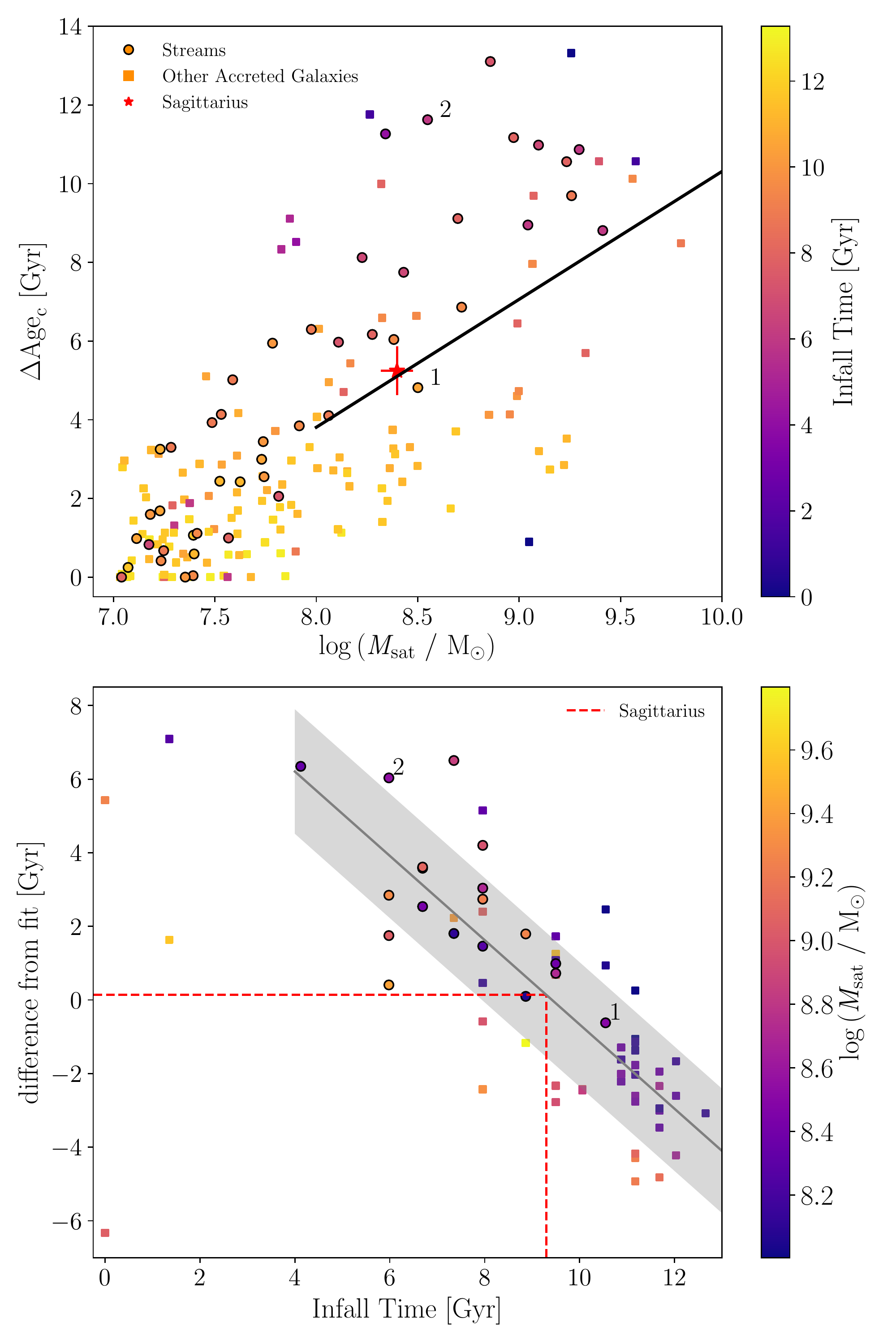}
    \caption{Top panel: the age range of the GCs which have been accreted with a satellite galaxy as a function of the parent galaxy's stellar mass, the solid black line represents the best fit line for satellite galaxies with a stellar mass greater than $10^8 \Msun$, the red symbol with error bar represents the position of the Sagittarius dwarf galaxy. Bottom panel: the difference from the line of best fit for each satellite galaxy above $10^8 \Msun$, the solid grey line represents the best fit line and the grey band represents the $1 \sigma$ scatter of the data around the fit, the red dotted line represents the method for estimating an infall time for the Sagittarius dwarf galaxy, discussed in Section \ref{6}. Each point represents an accreted galaxy, those accretion events that are seen as streams at $z=0$ are represented by circles and the rest of the accreted galaxies are represented by squares. The points labelled 1 and 2 will be used in Fig. \ref{fig:SFH.pdf} to investigate the star formation histories of two galaxies at the same mass but with different age ranges.}
    \label{fig:galaxymass_agerange_virialradius.pdf}
\end{figure}

\begin{figure}
	\includegraphics[width=\columnwidth]{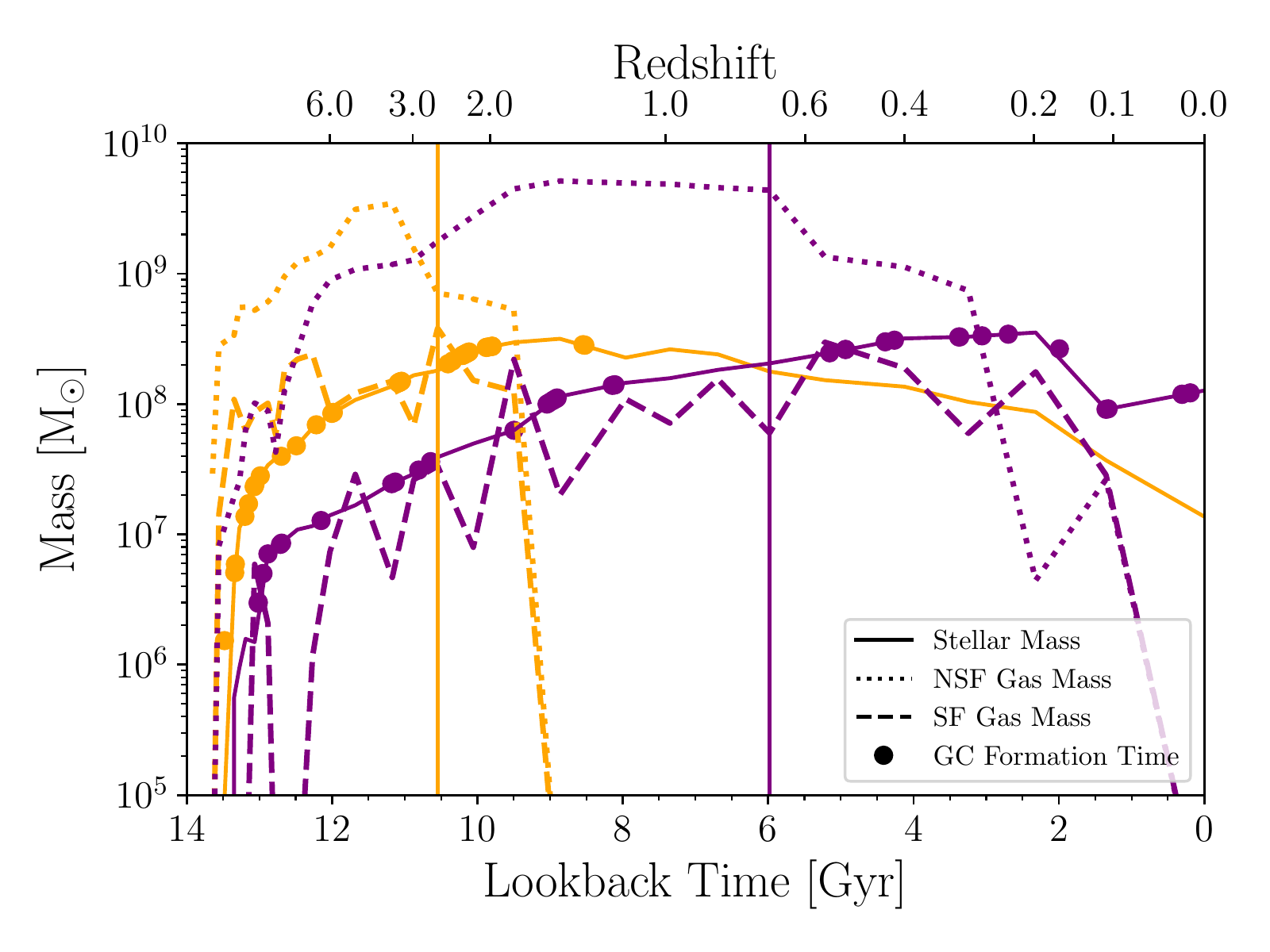}
    \caption{The star formation histories of two accreted galaxies that produce streams. The orange lines correspond to the galaxy labelled 1 in Fig. \ref{fig:galaxymass_agerange_virialradius.pdf} and the purple lines correspond to the galaxy labelled 2 in Fig. \ref{fig:galaxymass_agerange_virialradius.pdf}. The gas is split up into star forming (dashed lines) and non star forming (dotted lines). Note how the galaxy which crossed the virial radius (shown by the vertical lines) longer ago also stopped forming GCs longer ago, which is due to gas stripping.}
    \label{fig:SFH.pdf}
\end{figure}

The GC age range is a direct probe of the GC formation history: a greater GC age range indicates a more extended GC formation history. In the top panel of Fig. \ref{fig:galaxymass_agerange_virialradius.pdf} we see that, on average, more massive satellite galaxies have greater GC age ranges than lower mass satellite galaxies. In Fig. \ref{fig:galaxymass_agerange_virialradius.pdf} we separately show all galaxies that have been accreted, to assess whether the accretion events producing streams form a distinct group. Interestingly, the satellite galaxies that produce streams have a large GC age range for their mass. To understand this, we have to consider the reason why we see a stream -- the galaxy must have produced a stream-like structure as it fell into the main galaxy halo and the stars must have then stayed in this configuration for long enough for us to observe a stream at $z=0$. Therefore, a galaxy which causes an observable stream at present day is more likely to have fallen into the halo of the main galaxy more recently and has not had as long to disrupt. As we will discuss in Section \ref{5}, galaxies that entered the halo of the main galaxy more recently at a given mass have a greater GC age range, which would cause the streams to reside near the top of this distribution. \par 

Even though the relation is relatively tight, at a given galaxy mass, there is a large scatter in the GC age range -- up to 10 $\Gyr$ for the more massive satellites. We select two galaxies of approximately equal stellar mass but different GC age ranges, the two galaxies which are labelled as 1 and 2 in Fig. \ref{fig:galaxymass_agerange_virialradius.pdf}. We show the time evolution of their stellar and gas masses in Fig. \ref{fig:SFH.pdf}. The points on the line representing the stellar mass show the formation epochs of the GCs that survive untill present day (44 and 37 respectively).  In both cases, the mass of the gas and stellar component increases until the galaxy enters the halo of the main galaxy - shown by the vertical lines in Fig. \ref{fig:SFH.pdf}. Note here that we are limited by the snapshot time resolution of the simulation, so the fact that galaxy 1 starts to lose its non-star forming gas (NSF) before infall is not necessarily a real effect, but is infact because it entered the halo of the main galaxy at a time between the two snapshots. After infall, both galaxies start to lose NSF gas immediately and galaxy 1 also starts to lose its star forming (SF) gas. Galaxy 2 holds onto its SF gas for longer after infall, but in both cases we see a rapid and complete loss of all gas and a truncation in GC formation. Therefore, we see that at a fixed galaxy mass, the age range of the clusters associated with a satellite galaxy is potentially dependent on the infall time. Galaxy 1 has a smaller GC age range in Fig. \ref{fig:galaxymass_agerange_virialradius.pdf} than galaxy 2 because it fell into the halo of the main galaxy much earlier, shortening the GC formation history. \par

The low mass galaxies (i.e. $M_{\mathrm{sat}} < 10^{8} \Msun$) may have their GC formation truncated due to a variety of physical processes (such as stellar feedback), meaning that their infall time may not be well traced by their GC formation histories. To alleviate this, we do not include galaxies with masses lower than $10^8 \Msun$ in the rest of this analysis. We investigate the infall time being the reason for the scatter in the top panel of Fig. \ref{fig:galaxymass_agerange_virialradius.pdf} by subtracting off the mean relation of $\Delta \mathrm{Age_c}$ as a function of satellite galaxy stellar mass (solid line in top panel) and showing the residual against the infall time in the bottom panel of Fig. \ref{fig:galaxymass_agerange_virialradius.pdf}. We find that there is a strong correlation between difference from the line of best fit and infall time, indicating that the scatter in the age range at a given galaxy mass is indeed due to the satellite infall time. The galaxies that cross the virial radius of the main galaxy later build up their mass more slowly and have longer to form clusters free from severe environmental influences than those which build up their mass and fall into the halo of the main galaxy early in their evolution. This leads to a smaller cluster age range for satellites accreted early on. The fit to the data in the bottom panel is shown by the grey solid line. We do not include the four points with infall time less than $2 \Gyr$ ago, for two reasons. The satellite galaxy with a difference from fit of below $\rm{-6}$ is considered an outlier because it is a 'backsplash' galaxy \citep{Gill2005} i.e. it is an earlier crossing of the virial radius which causes this galaxy to stop forming GCs (this is discussed in section \ref{5.2}). The other three galaxies with infall time $< 2 \Gyr$ are outliers due their recent infall -- their $\Delta \rm{Age_c}$ is not yet fixed and could potentially continue to grow if the simulation was to continue running. \par

\subsection{GC formation after infall \label{5.2}}

\begin{figure}
	\includegraphics[width=\linewidth]{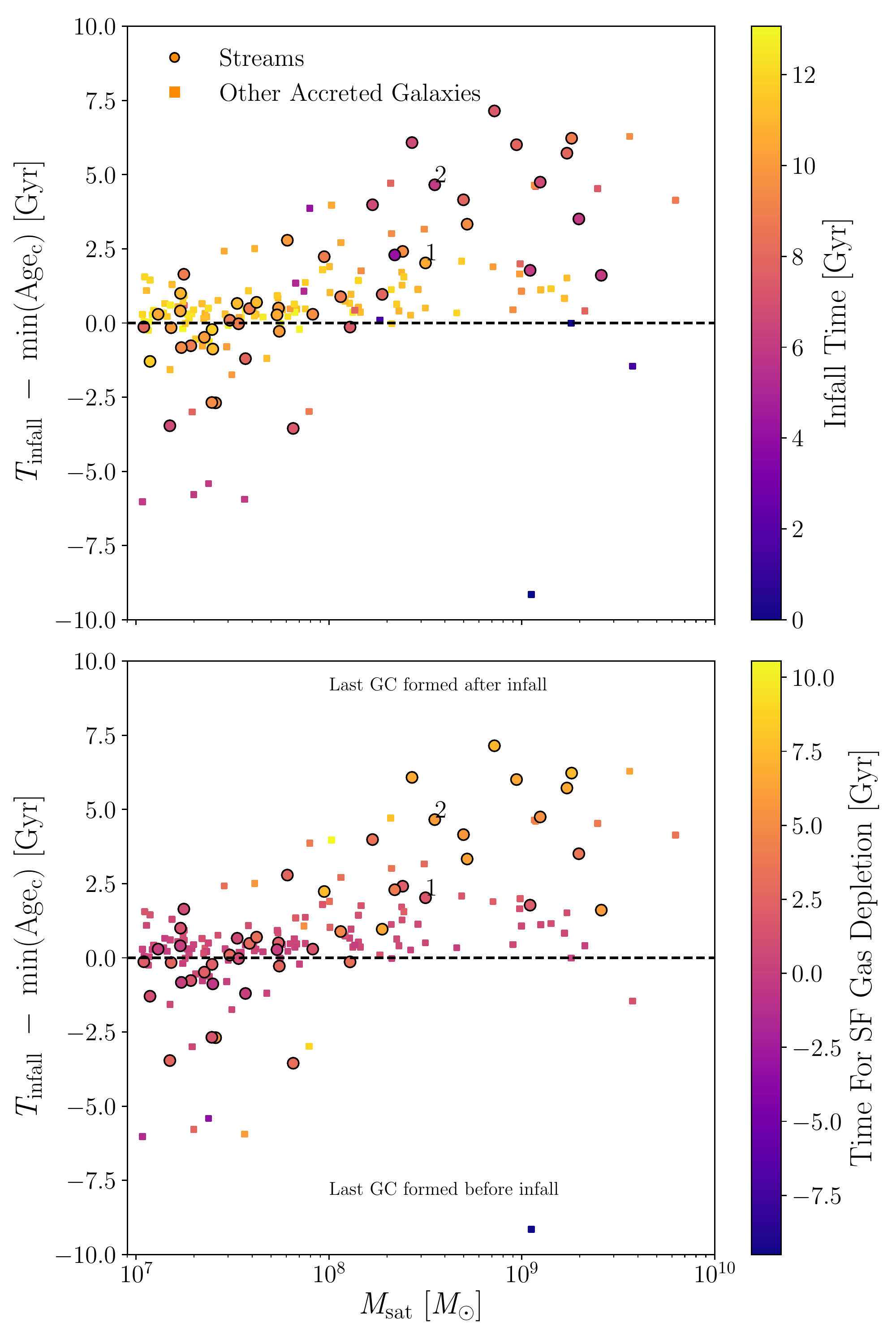}
    \caption{Time for which GCs continue to form after they have crossed the virial radius as a function of galaxy mass. The points are coloured by the infall time (top panel) and the time for which the galaxy retains its star forming gas after falling into the halo of the main galaxy (bottom panel) . }
    \label{fig:galaxymass_subplots_rvircross.pdf}
\end{figure}

\begin{figure}
	\includegraphics[width=\linewidth]{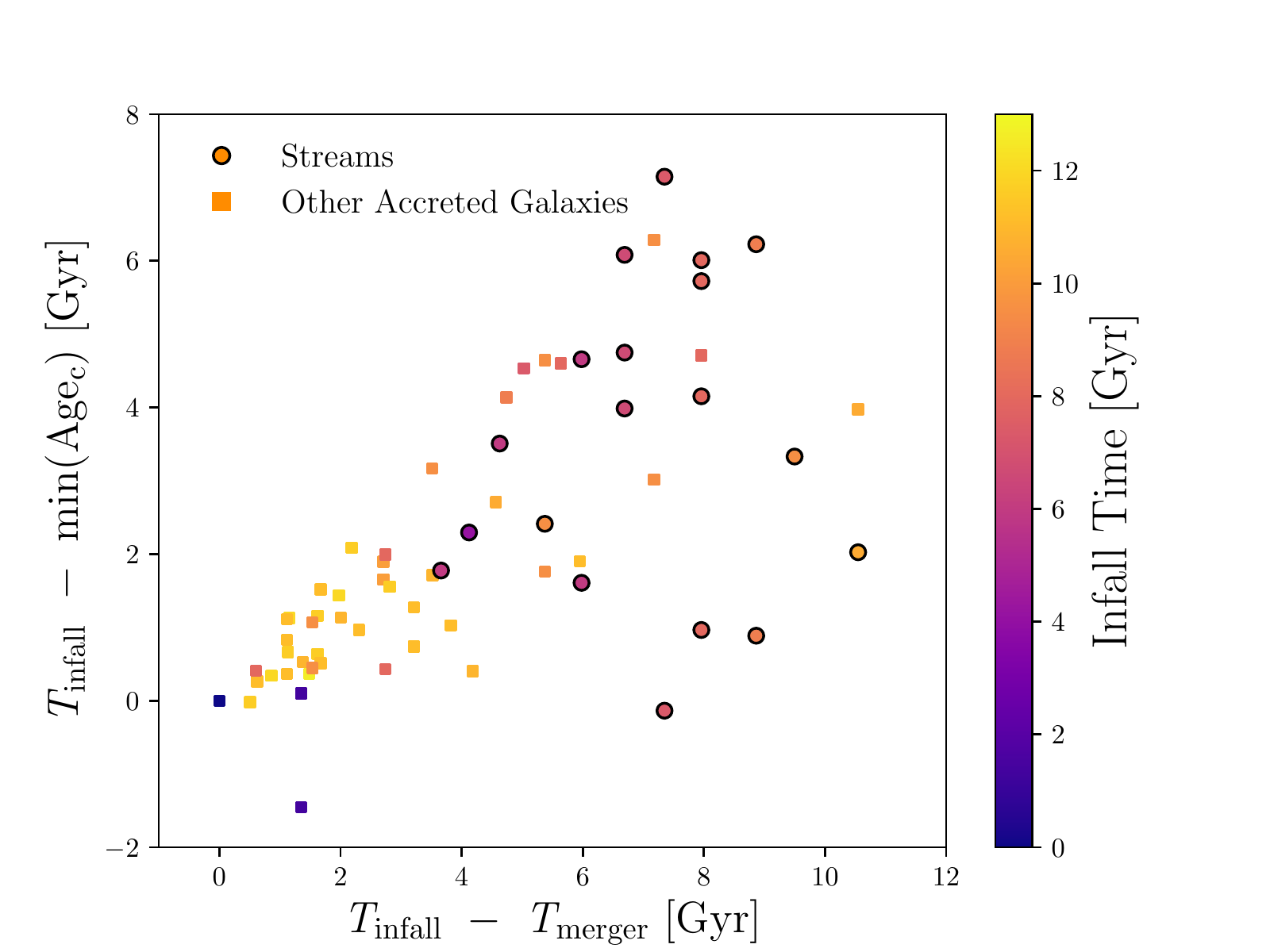}
    \caption{Time for which GCs continue to form after they have crossed the virial radius as a function of how long the satellite galaxy takes to completely merge with the main galaxy after it has crossed the virial radius. Only galaxies with $M > 10^8 \Msun$ are shown. The points are coloured by infall time. We see that faster mergers happen at earlier times.  }
    \label{fig:9.pdf}
\end{figure}

As discussed above, infall into the main galaxy halo and subsequent gas stripping is the main reason for the truncation of GC formation in satellite galaxies. However, some galaxies continue to form clusters after they have entered the halo of the main galaxy, we see this in the SMC and the LMC. We now investigate how long GCs continue to form after the satellite has fallen into the main group  ($\mathit{T}_\mathrm{infall} \mathrm{- min(Age_{c})}$) with respect to the galaxy mass and infall lookback time (Fig. \ref{fig:galaxymass_subplots_rvircross.pdf}).\par

We present the time for which GCs continue to form after the satellite galaxy has entered the halo of the main galaxy as a function of the satellite galaxy stellar mass in Fig. \ref{fig:galaxymass_subplots_rvircross.pdf} and we find that more massive galaxies can continue to form GCs for longer after entering the halo of the main galaxy. We can investigate this effect in relation to the time of infall (Fig. \ref{fig:galaxymass_subplots_rvircross.pdf}, top panel) and the time for which the galaxy retains its star forming gas after infall (Fig. \ref{fig:galaxymass_subplots_rvircross.pdf}, bottom panel). We will divide this discussion into whether the last GC forms during, after or before infall, that is when  $\mathit{T}_\mathrm{infall} \mathrm{- min(Age_{c})} \approx 0$,  $\gg0$ or $\ll0$ respectively. \par

Those satellites that stop forming GCs during infall are accreted early in the formation of the main galaxy, and lose their star forming gas almost immediately upon infall. In the early universe, when these galaxies are accreted, all halos are smaller. This means mergers happen on shorter time-scales and, consequently, star-forming gas gets stripped and GC formation truncates faster, which leads to a smaller GC age range after infall. 
This is demonstrated in Fig. \ref{fig:9.pdf}, where we present the time for which the satellite continues to form clusters after infall against the time it takes for the satellite to merge with the main galaxy after infall ($\mathit{T}_\mathrm{infall} -\mathit{T}_\mathrm{merger}$ ). We find that a quick truncation of GC formation after infall ($\mathit{T}_\mathrm{infall} \mathrm{- min(Age_{c})} \approx 0$) is due to a quick merger time and these quick mergers typically happen in early accretion events. Fig. \ref{fig:9.pdf} only shows satellite galaxies with a mass greater than $10^{8} \Msun$ because, as discussed above, below this mass some satellite galaxies stop forming GCs due to reasons other than infall into the main halo. This population of galaxies at $\mathit{T}_\mathrm{infall} \mathrm{- min(Age_{c})} \approx 0$ does not contain many stellar streams due to their early infall times, i.e. if a stream is produced, it is unlikely to survive until present day.\par

Those satellites that continue to form GCs after their infall are accreted later in the formation of the main galaxy. They show a dependency on their stellar mass. At greater masses, these galaxies can retain their star-forming gas for a longer time and retain high enough pressures to continue to form clusters. Many of the satellite galaxies in this population produce streams because the galaxies were accreted later and so the stream survives until present day.

Those satellites that stop forming GCs before their infall are accreted later in the formation of the main galaxy but stop forming clusters before they cross the central galaxy's virial radius. These are low mass galaxies ($\rm{M_{gal}} < 10^{8} \Msun$) that formed all of their GCs within a few $\Gyr$ (Fig. \ref{fig:galaxymass_agerange_virialradius.pdf} shows lower GC age ranges for lower mass galaxies). The low masses and densities of these galaxies imply that even the feedback from a burst of star formation can cause GC formation to cease. Many of these satellite galaxies also produce streams due to their later infall.\par

Finally, it is important to note here that we define infall time as the last time the satellite galaxy crossed the virial radius of the main galaxy. For most galaxies the last time they crossed the virial radius is an accurate representation of the interaction that caused the most change to the galaxy. However, in a few cases it is an earlier interaction with the main galaxy that causes the loss of star-forming gas and the truncation of GC formation -- these are known as backsplash galaxies \citep{Gill2005}. This affects the very blue point that has  $\mathit{T}_\mathrm{infall} \mathrm{- min(Age_{c})} \approx -9$   and a mass $\mathit{M}_\mathrm{gal} \approx \mathrm{10^{9}} \Msun$ in Fig. \ref{fig:galaxymass_subplots_rvircross.pdf}. It is an interaction with the main galaxy $9 \Gyr$ ago that causes this galaxy to lose star forming gas and stop forming GCs.

\section{Comparisons With Observations\label{6}}
 Observations of streams in the MW and other galaxies will be biased towards the most massive and recent accretion events, as these events are easier to observe both by overdensities of stars and kinematically.
 If we focus our sample on relatively high mass galaxies that were accreted recently, from Fig. \ref{fig:virialradius_met_streams.pdf} we find that they should host GC populations that are statistically younger, have a larger age range and are more metal-rich than the median across the entire accreted satellite population.\par
 
GCs on extragalactic stellar streams are much easier to study than individual stars, due to their higher surface brightness. Observations of the GC population outside 30 $\kpc$ of the centre of M31 have shown that a large fraction of these GCs are situated on streams \citep{Mackey2010} and this has also been found to be the case for other galaxies outside the Local Group (e.g. \citealt{Romanowsky2012,Powalka2018}).  As is the case for the Sagittarius dwarf, our simulations predict that the mean age of these GCs is younger than the other GCs associated with these galaxies. Age dating GCs at these distances (where colour-magnitude diagrams generally do not reach the main sequence turn-off) can be difficult. However, if these GCs are younger than $9-10 \Gyr$, they would not be expected to have an extended blue horizontal branch. Instead, they should have a compact red clump (or red horizontal branch) \citep[e.g.][]{Gratton2010}. Deep HST and/or ground based images will be able to test this prediction. In addition, relative ages between GC (sub)populations may be obtained by combining multi-band photometry with spectroscopy (Usher et al. in prep). With ages and metallicities of these GCs, parent galaxy mass and infall time could also be predicted for external galaxies.\par
 
 Throughout this work, we have compared various results to the GCs found in the Sagittarius dwarf galaxy, which is currently generating a large stellar stream in the halo of the MW. We show that the median metallicity and the median age of the clusters which have been associated with this stream are consistent with those found for the streams in this work at similar galaxy stellar masses. We can use Fig. \ref{fig:galaxymass_agerange_virialradius.pdf} to estimate the time at which the Sagittarius dwarf galaxy began its infall into the MW halo. Sagittarius was relatively massive before it fell into the halo of the MW with a stellar mass of  $\approx (2-3)\times 10^{8} \Msun$, (e.g. \citealt{Niederste-Ostholt2010}). Considering the GCs that have a high to moderate confidence of being associated with the Sagittarius stream from \cite{Forbes2010}\footnote{As in our previous analysis we exclude Berkley 29 and Whiting 1 based on their mass.} and the average ages from \cite{Forbes2010,Dotter2010,Dotter2011} and \cite{VandenBerg2013}, (see the compilation in Appendix~A of \citealt{Kruijssen2018b}) the GCs likely to be associated with the Sagittarius stream have an age range of $ 5.24 \Gyr$, shown in the top panel of Fig. \ref{fig:galaxymass_agerange_virialradius.pdf}. We can then find the difference of Sagittarius from the line of best fit, which can be used in the bottom panel of Fig. \ref{fig:galaxymass_agerange_virialradius.pdf} to estimate the infall time of Sagittarius -- shown by the red dotted line in this figure.  The uncertainty on the infall time is calculated by first considering the uncertainty on the difference of Sagittarius from the fit in the top panel. This includes the difference in the $\Delta \mathrm{Age_{c}}$ from the uncertainty on the stellar mass and the difference in the $\Delta \mathrm{Age_c}$ from the uncertainty on the age of the youngest and the oldest Sagittarius cluster. The uncertainty in the difference from fit is then propagated through to the bottom panel and is combined with the dispersion in the difference from fit against infall time parameter space to calculate a final uncertainty on the infall time.
We estimate an infall lookback time (time of virial radius crossing) of $9.3 \pm 1.8 \Gyr$.\citet{Dierickx2017} predict an infall lookback time of the Sagittarius dwarf of $8 \pm 1.5 \Gyr$ based on the age of the M giants in the stream calculated by \citet{Bellazzini2006}, which is consistent (albeit somewhat lower than) the value predicted by our analysis.

\section{Conclusions}
We present the GC properties of 15 MW-like haloes of the E-MOSAICS simulations. We specifically investigate the properties of  GCs that are associated with stellar streams relative to the rest of each galaxy's GC population. We find a large variation in the median ages and metallicities of the clusters on individual streams. It is found that more massive and recently accreted galaxies host GCs that are more metal rich and younger than the rest of the population, whereas less massive and earlier accreted galaxies harbour GCs that are older and more metal poor than the rest of the population. Applying this to M31, where massive and recent accretion events are easier to detect, we expect that GCs associated with stellar streams are, on average, younger that the rest of the population. This is consistent with observed GCs in M31 where GCs on streams are indeed found to be younger, on average, than GCs not on streams (Mackey et al. in prep.). \par

Two effects contribute to the GC age ranges of satellite galaxies. The first is that more massive streams host younger and more metal rich GCs because they entered the halo of the main galaxy more recently -- this allowed the satellites to continue form GCs for a longer time without being subject to strong environmental effects, resulting in a more extended GC formation history and younger GCs. Using the E-MOSAICS simulations, we find that the GC age range is more extended for more massive satellites, but there is a relatively large scatter at a given satellite mass. This scatter is determined by the infall time (i.e. the last time a galaxy enters the virial radius of the main galaxy, see Fig. \ref{fig:galaxymass_agerange_virialradius.pdf}). Galaxies that enter the halo of the main galaxy more recently have longer to evolve in isolation and therefore have a more extended GC formation history than galaxies of the same mass which entered the halo of the main galaxy early in cosmic history. The second effect is that more massive galaxies have more extended GC formation histories because they retain their star-forming gas for longer after infall into the main galaxy.\par

 With a reliable way of associating  observed GCs with stellar streams, it would be possible to take all of the GCs associated with a stellar stream and use their median metallicity and our relation between median GC metallicity and galaxy stellar mass shown in Fig. \ref{fig:virialradius_met_streams.pdf} to estimate a mass of their parent galaxy. Using this derived mass and the age range of the GCs, we could then place this galaxy in Fig. \ref{fig:galaxymass_agerange_virialradius.pdf} to estimate its infall time. Here this is done for Sagittarius and an infall lookback time of $9.3 \pm 1.8 \Gyr$ is calculated. \cite{Kruijssen2018b} predict the existence of three main satellites of the Milky Way, the least massive of which is Sagittarius. The other two satellites (the `Sausage' identified by \citealt{Myeong2018} and the enigmatic galaxy `Kraken' inferred by \citealt{Kruijssen2018b}) are indistinguishable in the age-metallicity relation of the Milky Way, but \citet{Kruijssen2018b} predict that they were accreted at $z < 2 $, i.e. more recently than $\approx 10 \Gyr$ ago. This suggests that all three of the major satellites of the Milky Way were accreted after $z=2$.\par
 Observations of GCs on streams are biased to massive stream progenitors, such as the Sagittarius stream, which explains why GCs observed to be on streams are younger on average than the rest of the GC population. The E-MOSAICS simulations show that when moving down to lower mass stream progenitor galaxies we can probe earlier accretion events, which contribute older GCs. However, to be able to probe these masses and infall times, better stellar stream detection and GC association methods are needed -- both of which will be facilitated within the Milky Way by current and future Gaia data releases.

\section*{Acknowledgements}
We thank the reviewer, Nicolas Martin, for a positive and helpful report. JP and NB gratefully acknowledge funding from a European Research Council consolidator grant (ERC-CoG-646928-Multi-Pop). JMDK gratefully acknowledges funding from the German Research Foundation (DFG) in the form of an Emmy Noether Research Group (grant number KR4801/1-1). JMDK and MRC gratefully acknowledge funding from the European Research Council (ERC) under the European Unions Horizon 2020 research and innovation programme via the ERC Starting Grant MUSTANG (grant number 714907). MRC is supported by a Fellowship from the International Max Planck Research School for Astronomy and Cosmic Physics at the University of Heidelberg (IMPRS-HD). NB and RAC are Royal Society University Research Fellows. This work used the DiRAC Data Centric system at Durham University, operated by the Institute for Computational Cosmology on behalf of the STFC DiRAC HPC Facility (www.dirac.ac.uk). This equipment was funded by BIS National E-infrastructure capital grant ST/K00042X/1, STFC capital grants ST/H008519/1 and ST/K00087X/1, STFC DiRAC Operations grant ST/K003267/1 and Durham University. DiRAC is part of the National E-Infrastructure. The study also made use of high performance computing facilities at Liverpool John Moores University, partly funded by the Royal Society and LJMU's Faculty of Engineering and Technology.



\bibliographystyle{mnras.bst}
\bibliography{sample}


\bsp	
\label{lastpage}
\end{document}